\documentstyle[12pt]{article}

\catcode`@=11

\def\beqra{\begin{eqnarray}} \def\eeqra{\end{eqnarray}}
\def\beqast{\begin{eqnarray*}} \def\eeqast{\end{eqnarray*}}
\def\beq{\begin{equation}}	\def\eeq{\end{equation}}
\def\be{\begin{enumerate}}   \def\ee{\end{enumerate}}

\def\gam{\gamma}
\def\Gam{\Gamma}

\def\eps{\epsilon}

\def\tha{\theta}

\def\mn{\mu\nu}

\def\ch{\@startsection{section}{1}{\z@}{-3ex plus-1ex minus-.2ex}%
	{2ex plus.2ex}{\large\sc{\normalsize}}}

\def\raisenot{\raise .5mm\hbox{/}}
\def\nota{\ \hbox{{$a$}\kern-.49em\hbox{/}}}
\def\notA{\hbox{{$A$}\kern-.54em\hbox{\raisenot}}}
\def\notb{\ \hbox{{$b$}\kern-.47em\hbox{/}}}
\def\notB{\ \hbox{{$B$}\kern-.60em\hbox{\raisenot}}}
\def\notc{\ \hbox{{$c$}\kern-.45em\hbox{/}}}
\def\notd{\ \hbox{{$d$}\kern-.53em\hbox{/}}}
\def\notbd{\ \hbox{{$D$}\kern-.61em\hbox{\raisenot}}} 
\def\note{\ \hbox{{$e$}\kern-.47em\hbox{/}}}
\def\notk{\ \hbox{{$k$}\kern-.51em\hbox{/}}}
\def\notp{\ \hbox{{$p$}\kern-.43em\hbox{/}}}
\def\notq{\ \hbox{{$q$}\kern-.47em\hbox{/}}}
\def\notW{\ \hbox{{$W$}\kern-.75em\hbox{\raisenot}}}
\def\notz{\ \hbox{{$Z$}\kern-.61em\hbox{\raisenot}}}

\def\notpa{\hbox{{$\partial$}\kern-.54em\hbox{\raisenot}}}

\def\fo{\hbox{{1}\kern-.25em\hbox{l}}}  

\def\and{\,{\rm and}\,}

\def\7#1#2{\mathop{\null#2}\limits^{#1}}	
\def\5#1#2{\mathop{\null#2}\limits_{#1}}	

\def\inbar{\vrule height1.5ex width.4pt depth0pt}
\def\IB{\relax{\rm I\kern-.18em B}}
\def\IC{\relax\leavevmode\hbox{\,$\inbar\kern-.3em{\rm C}$}}
\def\ID{\relax{\rm I\kern-.18em D}}
\def\IE{\relax{\rm I\kern-.18em E}}
\def\IF{\relax{\rm I\kern-.18em F}}
\def\IG{\relax\leavevmode\hbox{\,$\inbar\kern-.3em{\rm G}$}}
\def\IH{\relax{\rm I\kern-.18em H}}
\def\II{\relax{\rm I\kern-.18em I}}
\def\IK{\relax{\rm I\kern-.18em K}}
\def\IL{\relax{\rm I\kern-.18em L}}
\def\IM{\relax{\rm I\kern-.18em M}}
\def\IN{\relax{\rm I\kern-.18em N}}
\def\IO{\relax\leavevmode\hbox{\,$\inbar\kern-.3em{\rm O}$}}
\def\IP{\relax{\rm I\kern-.18em P}}
\def\IQ{\relax\leavevmode\hbox{\,$\inbar\kern-.3em{\rm Q}$}}
\def\IR{\relax{\rm I\kern-.18em R}}
\def\sed{\hbox{{\sf S}\kern-.4em\hbox{\sf S}}}
\def\ZZ{\relax{\sf Z\kern-.4em Z}}
\def\smIR{\hbox{{\footnotesize\rm I}\kern-.2em\hbox{\footnotesize\rm R}}}
\def\smIO{\ \hbox{{\footnotesize\rm I}\kern-.4em\hbox{\footnotesize\bf O}}}
\def\smIQ{\ \hbox{{\footnotesize\rm I}\kern-.5em\hbox{\footnotesize\bf Q}}}
\def\IGa{\relax{\rm I}\kern-.18em\Gamma}
\def\IPi{\relax{\rm I}\kern-.18em\Pi}
\def\IQt{\relax\leavevmode\hbox{$\kern.3em\inbar\kern-.3em\Theta$}}
\def\IOm{\relax\hbox{$\kern3.48pt\inbar\kern1.8pt\inbar\kern-5.28pt\Omega$}}

\def\ca#1{\relax\ifmmode {\cal#1} \else$\cal#1$\fi}	
\def\Sf#1{\relax\ifmmode\hbox{\sf#1}\else{\sf#1}\fi}	
\def\fibby{\ifcase\@ptsize 			
		\font\tenrm=cmfib8\or		
		\font\elvrm=cmfib8 scaled\magstephalf\or	
		\font\twlrm=cmfib8 scaled\magstep1 \fi}		
\def\TeXey{\ifcase\@ptsize\or\or 		
		\font\twlrm=cmr10 scaled\magstep1	
		\font\twlmi=cmmi10 scaled\magstep1	
		\font\twlit=cmti10 scaled\magstep1	
		\font\twlbf=cmbx10 scaled\magstep1\fi}	

\def\ch{\@startsection{section}{1}{\z@}{-3ex plus-1ex minus-.2ex}%
	{2ex plus.2ex}{\large\sc}}
\def\sch{\@startsection{subsection}{2}{\z@}{-1.5ex plus-1ex minus-.2ex}%
	{1pt plus.2ex}{\sc}}
\def\ssch{\@startsection{subsubsection}{3}{\z@}{-1ex plus-1ex minus-.2ex}%
     	{1pt plus.2ex}{\small\sc}}
\def\seceq{\@addtoreset{equation}{section}
	\def\theequation{\thesection.\arabic{equation}}}	

\def\con{\ifmmode \hbox{\bf*} \else{\bf*}\fi}	
\def\scon{\ifmmode \hbox{\footnotesize\rm\bf*} \else{\footnotesize\rm\bf*}\fi}

\def\0#1{\relax\ifmmode\mathaccent"7017{#1}
	\else\accent23#1\relax\fi}		

\def\haf{\frac{1}{2}}

\def\place#1#2#3{\vbox to0pt{\kern-\parskip\kern-7pt
                             \kern-#2truein\hbox{\kern#1truein #3}
                             \vss}\nointerlineskip}

\def\illustration #1 by #2 (#3){\vbox to #2{\hrule width #1 height 0pt depth
0pt
                                       \vfill\special{illustration #3}}}

\def\scaledillustration #1 by #2 (#3 scaled #4){{\dimen0=#1 \dimen1=#2
           \divide\dimen0 by 1000 \multiply\dimen0 by #4
            \divide\dimen1 by 1000 \multiply\dimen1 by #4
            \illustration \dimen0 by \dimen1 (#3 scaled #4)}}

\catcode`@=12

\begin{document}

\baselineskip=24pt

\hfill{DOE-ER-40757-021}

\hfill{CPP-93-21}

\hfill{June 22, 1994}

\begin{center}
 \Large{\bf  Dips in Partial Wave Amplitudes\\
from Final State Interactions}

\vspace{36pt}
{\normalsize Charles B. Chiu \ and\  Duane A. Dicus }

\vspace{16pt}

\end{center}

\begin{center}
{\bf Abstract}
\end{center}

\indent\indent We consider the dip-peak structures in the J=0 partial
wave amplitudes for processes $\gamma\gamma\rightarrow W^+W^-~
\mbox{and}~\gamma\gamma,gg\rightarrow t\overline{t}$ taking into account the
corresponding  Born term process and the rescattering process where the
intermediate state
is rescattered through the exchange of Higgs resonance state in the direct
channel.
\vfill

\pagebreak
\setcounter{page}{2}
\baselineskip=21pt

Near a direct channel resonance, final state interactions can be important even
for
relatively weakly interacting  particles.  Consider the production by photons
of a  pair
of $W$s, $\gam\gam \rightarrow W^+W^-$.  The Born amplitude is of order $e^2$
and
rescattering by the $W$s naively makes the amplitude of order $e^4$.  However,
if the
rescattering process corresponds to the exchange of, such as in this example,
the Higgs
in the direct channel, near the Higgs resonance the rescattering amplitude goes
as
$\frac{e^4m}{\Gam}\sim e^2$,  where $\Gam$ is the resonance width, and $m$ is
the
resonance mass.

A characteristic signature of a final state interaction near a resonance
is a {\it dip} in  the scattering amplitude.[1-7]
This dip occurs for the
part of the amplitude where the rescattering particles are on mass shell and
when there
is only a single channel involved.\footnote{See in particular the discussion in
the
Appendix of Ref. 5.}  In  practice,the dip may be washed out by the part of the
amplitude with the particles off-shell, other particles produced and
rescattered into the
desired final state (including other polarizations of the final state
particles), and even
contributions from other particles than the final state particles to
$\Gamma$. Our goal in this  note is to consider a few simple
reactions in the Standard Model and check to what
extent a dip will manifest itself in the partial wave amplitude. We choose to
work
directly with the J=0 partial wave amplitude, since the interference effect
should show up
most prominently in partial wave amplitudes.

As a definite, and uncomplicated example consider  again $\gam\gam\rightarrow
W^+W^-$.
First take only the Goldstone boson part of the $W$s, $\gam\gam\rightarrow
\chi^+\chi^-$
and work in the $\xi=1$ gauge where $m_\chi=M_W$.  The Born $J=0$ amplitude is
\begin{equation}
a_{++}^{{\rm Born}}= \frac{e^2}{16\pi}\;\frac{(1-\beta^2)}{\beta}\;
\ln\;\frac{1+\beta}{1-\beta}
\end{equation}
where $\beta$ is the $\chi$ velocity, $\beta=(1-M_W^2/E^2)^{\haf}$, with $E$
the initial
photon energy.

The amplitude for two photons to produce an off-shell Higgs is
\begin{eqnarray}
\eps_\mu(k_1)\eps_\nu (k_2) T^{\mn} &=&
\frac{e^3M_W\pi^2}{\sin\tha_W(2\pi)^4}\;\left(g^{\mn}
- \frac{2k_1^\nu k_2^\mu}{s}\right) \eps_\mu(k_1)\eps_\nu(k_2) \nonumber
\\[4pt]
&&\times \left[ \frac{m^2_H}{M^2_W} + \frac{m_H^2}{s}\; I \right]
\end{eqnarray}
where $s=4E^2$ and
\begin{eqnarray}
I=\left[ \ln\; \frac{1+\beta}{1-\beta} - i\pi\right]^2
\end{eqnarray}
for $s>4M^2_W$.  Thus the contribution to $\gam\gam\rightarrow\chi^+\chi^-$
from
rescattering of the $\chi$ is
\begin{equation}
\frac{e^4\pi m_H^2}{32\sin^2\tha_W(2\pi)^4}\;\left[\frac{m_H^2}{M_W^2} +
\frac{m_H^2}{s}\,I\right]\; \frac{1}{s-m_H^2+im_H\Gam}\,.
\end{equation}
If we take $\Gam$ in (4) to be only the $H\rightarrow \chi\chi$ contribution
\begin{equation}
\Gam_{H\rightarrow \chi\chi} = \frac{e^2m_H^3}{64\;\pi M_W^2\sin^2\tha_W}\;
\left[
1-\frac{4M_W^2}{m_H^2} \right]^{1/2}
\end{equation}
then the contribution to the $J=0$ partial wave amplitude from the imaginary
part of $I$
exactly cancels $a^{{\rm Born}}_{++}$ given by (1).  Hence the dip of curve-b
as shown in Figure 1.
But even in this idealized example we should consider as well the real part of
the square
bracket in (4).  This is  shown as the curve-c in Fig. 1, where there is a dip
-- peak structure --
the Higgs resonance asserts itself at energies slightly above the Higgs mass.
The dip is
not an exact zero if we use (5); however if we replace $m_H\Gamma$ in (4) by an
energy
dependent expression $\sqrt s\Gamma(s)$, the net effect of which is simply to
replace
$m_H^2$ by $s$ only in the radical of (5), the dip is an exact zero. Its
position can be
calculated from the equations above - surprisingly it has only a very weak
dependence on
$m_H$; for $m_H$ equal 400, 600, 800 and 1000 GeV the zero occurs at $\sqrt s$
of 390,
510, 570 and 560 GeV which means there are very broad valleys for the larger
$m_H$.
Because our results are only strictly correct at $\sqrt s$ near $m_H$, we will
ignore
energy dependence in the widths in the remainder of this paper.

A more realistic example is to consider the scattering into longitudinal $W$s,
$a_{++;LL}$, where the Born amplitude is still given by (1).  But, of course,
we must
allow all polarizations of the $W$s in the loop intergrals connecting
$\gam\gam$ to $H$.
This has the effect of replacing the square bracket in (2) and (4)
by\footnote{In Ref. 6,
the interference effects between resonant and nonresonant amplitude for
$\gamma\gamma
\rightarrow W^+W^-$ process is also investigated.  Our expressions of this
process agree
with their expressions at resonance energy.  But there are differences at
off-shell values
due to the fact that some $m^2_H$ factors in our expressions are replaced by
$s$ in
theirs. For an off-shell Higgs the expression for the loop given in Eq. (6) is
not unique.
These differences presumably reflect a different choice of gauge. We work in
the $R_{\xi}$
gauge where the couplings of the Goldstone bosons depend on $m_H$; in the
unitary gauge,
for example, $m_H$ does not appear. }
 \begin{equation} 6+\frac{m^2_H}{M^2_W}
+ \left(\frac{m_H^2}{s}+\frac{6M_W^2}{s}-4\right) \;I\,. \end{equation} Also we
must use
the full $H$ width in the denominator of (4), $\Gam_{H\rightarrow WW}+
\Gam_{H\rightarrow
ZZ}$ (we are ignoring the top quark at this point in the discussion)  and
multiply (4) by
$- s(1+\beta^2)/2m_H^2$ for the longitudinal polarization vectors.  This is
shown as the
curve-b in Fig. 2.  Although there is a vague  resemblance in the overall
dip-peak
structures for the $\gamma\gamma\rightarrow \chi\chi$ the dip has disappeared
completely
in $\gamma\gamma\rightarrow W_LW_L$.

Now include the top quark by including it in the Higg's width
\begin{equation}
\Gamma_{H\rightarrow t\overline{t}}=\frac{e^2 m^2_tm_HN_c}{32\pi M_W^2
sin^2\theta_W}\;
\left[1-\frac{4m^2_t}{m^2_H}\right]^{3/2}
\end{equation}
and including it in the loop which connects $\gamma\gamma$ to $H$.  This means
replacing
the square bracket in (2) and (4) not just by (6) but by (6) plus
\begin{equation}
-4e^2_tN_c\;\frac{m^2_t}{M_W^2}\; \left[ 1+\left(\frac{ m^2_t}{s}
-\frac{1}{4}\right) I_t
\right]
\end{equation}
where the charge of the top quark is $e_t=2/3$, the number of colors, $N_c$, is
3 and
$I_t$ is given by (3) with $\beta$ replaced by
$\beta_t=\left(1-\frac{m^2_t}{E^2}\right)
^{1/2}$ if $s>4m^2_t$ or, if $s<4m^2_t$,
\begin{equation}
I_t=-4\left[\sin^{-1}\left(\frac{E}{m_t}\right)\right]^2\,.
\end{equation}
The effect of the top is shown in Fig. 3. The dip has reappeared for all $m_t$
values as
long as $m_H$ is not too large.[6]

Notice that beyond the tree level ,
diagrams which have been included are only those which involve one-triangular
loop
followed by the rescattering through the exchange of the direct channel Higgs
resonance.
This is a reasonable approximation near the resonance region, where the
resonance
contribution is dominant. On the other hand, in the two ``tail" regions of the
resonance,
where the resonance contribution is no longer dominating, there are competing
diagrams
which have been omitted and our predictions are expected to be less
reliable.

For transverse $Ws$ the Born term is slightly different than (1)
\begin{eqnarray}
a^{{\rm Born},J=0}_{++;++} &=& \frac{e^2}{16\pi}\;\frac{(1+\beta)^2}{\beta}\;
\ln\;
\frac{1+\beta}{1-\beta} \\[5pt]
a^{{\rm Born},J=0}_{++;--} &=& \frac{e^2}{16\pi}\;\frac{(1-\beta)^2}{\beta}\;
\ln\;
\frac{1+\beta}{1-\beta}\,.
\end{eqnarray}
Thus the relative magnitudes of the $J=0$ Born terms for $++:LL:--$ polarized
final $W$'s
are 1~: $\left(\frac{1-\beta}{1+\beta}\right)
:\;\left(\frac{1-\beta}{1+\beta}\right)^2$
while the final state Higgs interaction for these polarizations goes as $1\;:$
$\frac{s(1+\beta^2)}{4M^2_W}$ :1. For the $++$ final state, the ratio of the
Born term
to the loop contribution compared to the $LL$ final state case, is enhanced by
a
factor of $\left(\frac{1+\beta}{1-\beta}\right) \cdot
\frac{s(1+\beta^2)}{4M^2_W}$.  This
factor is about 260 for $s=(400{\rm GeV})^2$.  Thus the Born term for the $++$
polarization dominates over the corresponding loop contribution.  No effect of
Higgs
interactions can be seen.
 Conversely the $--$ amplitude is very small as shown in Fig. 4. Notice that
the
coordinate is in the logorithmic scale here.  It shows a full dip-peak
structure with the
exact behavior depending on the values of $m_t$, which are taken to be 140, 170
and 200 GeV for curves (a), (b) and (c) respectively.

Consider now the production of top quarks by photons or gluons,
$\gamma\gamma\rightarrow
t\overline{t}$ and $gg \rightarrow t\overline{t}$.  Again we can have a final
state
interaction where the photons or gluons couple to the Higgs through a loop and
the Higgs
couples directly to the final tops.  For photons the $J=0$ Born terms for the
two
independent helicity amplitudes are given by
\begin{eqnarray}
a^{{\rm Born},J=0}_{++;\haf\haf} &=&\frac{e^2e^2_t}{16\pi} (1-\beta^2)^{1/2}
\;\frac{(1+\beta)}{\beta}\;\ln\; \frac{1+\beta}{1-\beta} \\[5pt]
a^{{\rm Born},J=0}_{++;-\haf-\haf} &=& - \frac{e^2e^2_t}{16\pi}
(1-\beta^2)^{1/2}
\;\frac{1-\beta}{\beta}\;
\ln\;
\frac{1+\beta}{1-\beta}\,.
\end{eqnarray}
where $\beta $ was called $\beta_t$ above.  For gluons of  equal color the
$J=0$ Born
terms are also given by (12) and (13) except for some different coupling
factors.  The
Higgs interaction is different in the two cases because we can have different
particles
in the loop.  For photons it is given by (4) with (6) plus (8) in the square
bracket; for
gluons only (8) is in the square bracket (in order to compare the effect of
different
contributions to the loop we will ignore again the different gluon coupling
factors).
For both photons and gluons Eq. (4) must also be multipled by
$$2m_tE\beta_t /m_H^2 $$
for the spinors of the final tops.

 The results of $|a_{++;\haf\haf}|^2$ for top mass of 140, 170, or 200 GeV
 are shown in Figs. 5a, 5b and 5c respectively. In each figure the Higgs mass
is taken to be
 500, 600, or 800 GeV. The $gg$ process has big dips with small peaks (curves
b, d and f), while the
$\gamma\gamma$ process has the reverse (curves a, c and e).  Because (12) and
(13) have opposite sign this is
reversed for  $|a_{++;-\haf-\haf}|^2$ as shown in Fig. 6. Here the photon
process has a
small dip (curves a and c) which becomes a peak if $m_H$ is large (e.g. curve
e).
The gluon process has a large peak (curves b, d and f).
Since the $\haf\haf$  amplitude is much larger than the $-\haf-\haf$ amplitude
the total
cross section for $gg\rightarrow t\overline{t}$ will have a dip, except for
$m_H$ very near
threshold[7].  Similar results hold for other top masses.

A final word of caution is in order here. We have
discussed the behavior of the square of the J=0 partial wave amplitude for
various
processes. As alluded to earlier, the partial wave amplitude behavior gives the
 most
prominent display of the interference effect. On the other hand, at the cross
section
level, due to the contribution of the Born term to other partial waves, the
magnitude of the interference effect compared to that at the partial wave
amplitude
level is expected to be greatly reduced. We refer the reader to, for examples
Refs 4, 6 and
7, for typical suppressions involved. Reference 7 seems to show that th
struture is potentially
observable at CERN Large Hadron Collider.

In conclusion we find that  near
the Higgs resonance region, the dip structure which appears in one piece of the
Goldstone
boson production contribution shows up as more complex dip-peak patterns in the
full
partial wave amplitude and various different final helicity states and for
different intermediate state contributions.  Photon- and gluon-$t\overline{t}$
production have opposite dip-peak structure with the gluon process having
mostly dips for
the dominant helicity state.

Much of this work was done in collaboration with A. Stange and S. Willenbrock
and we
are very grateful for their help. This work was supported in part by the U. S.
Department
of Energy under grant DE-FG03-93-ER40757.

\vspace{0.25in}

\centerline{\bf References}

\begin{description}
\item[1.] L. Resnick, Phys. Rev. {\bf D2}, 1975 (1970); J. Pumplin, Phys. Rev.
{\bf D2},
1859 (1970); T. Bauer, Phys. Rev. Lett. {\bf 25}, 485 (1970).
\item[2.] J.-L. Basdevant and E. Berger, Phys. Rev. {\bf D16}, 657(1977); {\bf
D19}, 239
(1979); {\bf D19}, 246 (1979).
\item[3.] D. Morgan and M. Pennington, Z. Phys. {\bf C37}, 431 (1988).
\item[4.] K. Gaemers and F. Hoogeveen, Phys. Lett. {\bf 146B}, 347(1984).
\item[5.] J. L. Basdevant, E. Berger, D. Dicus, C. Kao and
S. Willenbrock, Phys. Lett. {\bf 313B}, 402 (1993).
\item[6.] D. A. Morris, D. N. Truong
and D. Zappala,``Higgs boson interference in $\gamma\gamma\rightarrow W^+W^-$",
UCLA
preprint TEP/93/35, Oct 1993.
\item[7.] D. Dicus, A. Stange and S.
Willenbrock, ``Higgs Decay to Top Quarks at Hadron Colliders", Illinois
preprint
ILL-(TH)-94-9 (1994).
 \end{description}

\pagebreak

\centerline{\bf Figure Captions}

\begin{description}

\item[Figure 1.]  $|a_{++;LL}|^2$ vs. $E$ for $\gamma\gamma\rightarrow$
Goldstone
bosons, where $E$ is the center of mass energy of one photon.  The Higgs mass
is taken to be 600  GeV. The solid line (a) is the Born
contribution.  The short-long dashed line (b) is
$\gamma\gamma\rightarrow
\chi^+\chi^-$ including only the imaginary part of the $\chi$ loop.  The short
dashed line
(c) includes all of the $\chi$ loop.

\item[Figure 2.]  $|a_{++;LL}|^2$ vs. $E$ for $\gamma\gamma\rightarrow W_LW_L$
ignoring
top quarks.  $m_H$ equals 600 GeV. The solid and short dashed lines, (a) and
(c) are as in
Fig.\ 1 for the $\gamma\gamma\rightarrow \chi^+\chi^-$ process .
The short-long dashed line (b) is the process
$\gamma\gamma\rightarrow W^*W^*\rightarrow H\rightarrow W_LW_L$ including all
polarizations of $Ws$ in the loop connecting the photons to the Higgs and the
full Higgs
width in the propagator (neglecting top quarks).

\item[Figure 3.] $|a_{++;LL}|^2$ vs. $E$ for  the physical process
$\gamma\gamma\rightarrow W_LW_L$ including the total gauge contribution (Fig.\
2) plus the
top quark contribution.  The solid line is the Born term as in Fig.\ 1.  The
short-long dashed
line (a), short dashed line (b), and long dashed line (c) are for $m_t=140$,
170, and 200 GeV.
Fig.\ 3a has $m_H=500$ GeV, Fig.\ 3b has $m_H=600$ GeV and Fig.\ 3c has
$m_H=800$ GeV.

\item[Figure 4.]  $|a_{++;--}|^2$ vs. $E$ for $\gamma\gamma\rightarrow W_TW_T $
the
Higgs Mass is 500 GeV. The solid line is the Born amplitude.  The other lines,
(a),
(b), (c), include the full gauge contribution to the final state interaction as
well as a
top quark of mass 140, 170, and 200 GeV as in Fig.\ 3.

\item[Figure 5.]  $|a_{++;\haf\haf}|^2$ vs. $E$ for $\gamma\gamma\rightarrow
t\overline{t}$  or $gg\rightarrow t\overline{t}$.  The solid
line is the Born term which has been normalized to be the same for the gluon
and photon
processes. Fig.\ 5a has $m_t=140$ GeV, Fig.\ 5b has $m_t=170$ GeV and Fig.\ 5c
has $m_t=200$ GeV.
In each figure the Higgs mass is taken to be 500, 600, and 800 GeV,
the corresponding curves (a), (c) and (e) are for the photon process, while the
corresponding
curves (b), (d) and (f) are for the gluon process.

\item[Figure 6.]  $|a_{++;-\haf-\haf}|^2$ vs. $E$ for $m_t = 170$ GeV.
Otherwise the same as Fig.\ 5.

\end{description}

\end{document}